\documentstyle[aps,multicol,epsf,epsfig,psfrag,amssymb]{revtex}
\newcommand{\beq}{\begin{equation}}
\newcommand{\eeq}{\end{equation}}
\newcommand{\bea}{\begin{eqnarray}}
\newcommand{\eea}{\end{eqnarray}}
\def\al{\alpha}
\def\be{\beta}
\def\ga{\gamma}
\def\de{\delta}
\def\si{\sigma}
\def\C{{\cal{C}}}
\def\O{{\cal{O}}}
\def\wt{\widetilde}
\def\ol{\overline}
\def\l{\left}
\def\r{\right}
\def\nl{\nonumber\\}
\begin{document}
\draft
\preprint{\begin{tabular}{l}
\hbox to\hsize{July, 1999 \hfill KAIST-99/04}\\[-3mm]
\hbox to\hsize{           \hfill hep-ph/9611369 }\\[5mm] \end{tabular} }
\title{Fully supersymmetric CP violations \\ in the kaon system } 
\author{S. Baek, J.-H. Jang, P. Ko and J. H. Park \\}
\address{
Department of Physics, KAIST \\  
Taejon 305-701, Korea }
\maketitle
\thispagestyle{empty}

\begin{abstract}
We show that, on the contrary to the usual claims, fully supersymmetric 
CP violations in the kaon system are possible through the gluino 
mediated flavor changing interactions. Both $\epsilon_K$ and ${\rm Re} 
(\epsilon' / \epsilon_K)$ can be accommodated for relatively 
large $\tan\beta$ without any fine tunings or contradictions 
to the FCNC and EDM constraints. 
\end{abstract}

\pacs{PACS numbers:}

\begin{multicols}{2}

Until this year, the only CP violation observed was in $K_{L} \rightarrow 
2 \pi$ \cite{cronin}, which could be attributed to $\Delta S = 2$ $K^0 - 
\overline{K^0}$ mixing. 
The mixing parameter $\epsilon_K$ is accurately measured by now :
$\epsilon_K = e^{i \pi/4} ~( 2.280 \pm 0.013 ) \times 10^{-3}$ \cite{pdg}. 
Recent observation of ${\rm Re} (\epsilon' / \epsilon_K )$ by KTeV 
collaboration, 
$
{\rm Re} ( \epsilon' / \epsilon_K ) = (28 \pm 4) \times 10^{-4}
$
\cite{ktev}, 
nicely confirms the earlier NA31 experiment \cite{na31}
$
{\rm Re} ( \epsilon' / \epsilon_K ) = (23 \pm 7) \times 10^{-4}.
$
This nonvanishing number indicates unambiguously the existence of  CP 
violation in the decay amplitude $(\Delta S = 1)$.
These two parameters quantifying CP violations in the kaon system can be 
accommodated by the KM phase in the Glashow-Salam-Weinberg's standard 
model (SM). The SM prediction for the latter is about $5 \times 10^{-4}$ and 
lies in the lower side of the data, although theoretical uncertainties from 
nonperturbative matrix elements and the strange quark mass are rather 
large \cite{buras_1}.  

However, it would be interesting to consider a possibility that these CP 
violations have their origin entirely different from the KM phase 
in the SM, in particular 
in the framework of various extensions of the SM including supersymmetric 
models \cite{recent}. In the minimal supersymmetric standard model 
(MSSM) considered in this work, there are many new CP violating phases that 
fall into two categories : phases with flavor 
preserving (FP) and flavor changing (FC), each of which is constrained by 
electron/neutron electric dipole moments (EDM's) and the $\epsilon_K$,
respectively. Recently, it was shown that the FP and CP 
violating phases in $A_t$ and $\mu$ in the more minimal SUSY model do not 
generate enough $\epsilon_K$ \cite{abel} \cite{baek2} or new phase shift in 
$B^0 - \overline{B^0}$ \cite{abel} \cite{baek1}, 
although they can lead to a large 
direct CP asymmetry in $B\rightarrow X_s \gamma$ upto $\sim\pm 16\%$ if 
charginos and stops are light enough \cite{baek1}. However, another class 
of CP violating phases in the flavor changing quark-squark-gluino vertices 
in the MSSM may be relevant to CP violations in the $K$ meson system. 

In this letter, we show that all the observed CP violating phenomena in the 
kaon system
in fact can be accommodated in terms of a single complex number 
$( \delta_{12}^d )_{LL}$ that parameterizes the squark mass mixings in the 
chirality and flavor spaces 
for relatively large $\tan\beta$ without any fine tuning or any 
contradictions with experimental data on FCNC, even if  $\delta_{\rm KM}=0$. 
We assume that CKM matrix is real in the most of this letter for simplicity 
and maximizing the effect of our mechanism. 
The case where the KM phase is nonzero is discussed in brief, and the details
will be given elsewhere \cite{future}. 

In order to study the gluino (photino) mediated flavor changing phenomena 
in the quark (lepton) sector
such as $\Delta m_K, \epsilon_K$ and ${\rm Re} ( \epsilon' / \epsilon_K )$ or 
lepton flavor violations, 
it is convenient to use the so-called mass insertion approximation (MIA) 
\cite{mia}. 
The quark-squark-gluino vertex is flavor diagonal in the MIA, and
the flavor/chirality mixing occur through the insertion of 
$( \delta_{ij}^d )_{AB}$, where $i,j=1,2,3$ and $A,B=L,R$ denote the flavors 
of the squark under consideration and the chiralities of its superpartner. 
The superscript denotes that the down type squark mass matrix is involved.
The parameters $( \delta_{ij}^d )_{AB}$ characterize the size of the 
gluino-mediated flavor changing amplitudes, and they may be CP violating  
complex numbers, in general. 
In the following, diagrams involving charged 
Higgs, chargino and neutralino will be ignored, since they are suppressed by 
$\alpha_w /\alpha_s$ compare to the gluino-squark loops unless 
gluino/squarks are very heavy. 
This should be a good starting point for studying the SUSY FCNC/CP problems.

Now, if one saturates $\Delta m_K$ and $\epsilon_K$ with 
$( \delta_{12}^d )_{LL}$ alone, the resulting ${\rm Re} ( \epsilon^{'} / 
\epsilon_K )$ is too small by more than an order of magnitude,
unless one invokes some finetuning \cite{gabriel}.   
Recently, Masiero and Murayama showed that this conclusion can be evaded 
in generalized SUSY models 
\cite{mm} with a few reasonable assumptions on the size of the 
$(\delta_{12}^d)_{LR}$ and the relations between the down quark Yukawa 
couplings and the CKM mixings. But they did not consider possibility to 
generate $\epsilon_K$ from a supersymmetric phase in Ref.~\cite{mm}, and also
predict too large neutron EDM which is very close to the current upper limit.

In the following, we show that there is another {\it generic} way 
to saturate $\epsilon^{'}/\epsilon_K$ in supersymmetric models 
if $|\mu \tan\beta|$ is relatively large, say  $\sim 10 - 20$ TeV. 
Moreover, both $\epsilon_K$ and 
${\rm Re} ( \epsilon^{'} / \epsilon_K )$ can be generated by a single 
CP violating complex parameter in the MSSM. In other words, fully 
supersymmetric CP violations are possible in the kaon system. 
The argument goes as follows : 
if $| ( \delta_{12}^d )_{LL} | \sim O(10^{-3} - 10^{-2})$ with the phase 
$\sim O(1)$ saturates $\epsilon_K$, this same parameter can
lead to a sizable ${\rm Re} (\epsilon' / \epsilon_K )$ through the 
$( \delta_{12}^d )_{LL}$ insertion followed by the FP $(LR)$ mass 
insertion, which is proportional to 
\[
(\delta_{22}^d)_{LR} \equiv m_s ( A_s^* - \mu \tan\beta) / \tilde{m}^2 
\sim O(10^{-2}), 
\]
where $\tilde{m} $ denotes the common squark mass in the MIA.
It should be emphasized that the induced $( \delta_{12}^d )_{LR}^{\rm ind}
\equiv ( \delta_{12}^d )_{LL} \times (\delta_{22}^d)_{LR} $ 
is different from the conventional  $ ( \delta_{12}^d )_{LR}$ in the 
literature. The loop functions for these two $LR$ insertions  are different 
with each other in general. The $LR$ mixing $( \delta_{12} )_{LR}^{\rm ind}$ 
induced by $( \delta_{12}^d )_{LL}$  is typically very small in size 
$\sim O(10^{-5})$, but this is enough to generate the full size of 
${\rm Re} (\epsilon' / \epsilon_K)$ as shown below. 
Our spirit to generate supersymmetric ${\rm Re} 
(\epsilon^{'} / \epsilon_K)$ is different from  Ref.~\cite{mm}, 
where the $LR$ mass matrix form is assumed to be similar to
the Yukawa matrix so that they predict the neutron EDM to be close to the 
current upper limit. On the other hand, our model does not suffer from the
EDM constraint at all, as shown below.  

Let us first consider the gluino-squark contributions to the 
$K^0 - \overline{K^0}$ mixing due to two insertions of 
$( \delta_{12}^d )_{LL}$. The corresponding $\Delta S = 2$ effective 
Hamiltonian is given by 
\[
{\cal H}_{\rm eff} (\Delta S =2) 
= C_{1} \ol{d}_L^\al \ga_\mu s_L^\al ~\ol{d}_L^\be \ga^\mu s_L^\be
\]
with the Wilson coefficient $C_{1}$ being 
\bea
  C_1 = -{\al_s^2 \over 216 \wt{m}^2} 
         (\de^d_{12})^2_{LL} \left[ 24 x f_6(x) + 66 \wt{f}_6(x) \right] .
\eea
Here, $x=m^2_{\wt{g}}/\tilde{m}^2$ and the loop functions $f_6(x)$ and 
$\wt{f}_6(x)$ are given in~\cite{gabbiani:96}.
The double mass insertion diagrams of FC $LL$ followed by the FP $LR$ 
generate another operators : 
$ \wt{Q}_2 = \ol{d}_L^\al  s_R^\al \ol{d}_L^\be s_R^\be$ and 
$ \wt{Q}_3 = \ol{d}_L^\al  s_R^\be \ol{d}_L^\be s_R^\al$, whose 
Wilson coefficients 
are proportional to $\left[ (\de_{12}^d)_{LR}^{\rm ind} \right]^2$.
Since their effects on $\Delta m_K$ and $\epsilon_K$ are negligible, we do 
not show them here explicitly although their effects have been included in 
the numerical analyses. 

Now we turn to the $\Delta S=1$ effective Hamiltonian 
${\cal H}_{\rm eff} ( \Delta S = 1 ) = \sum_{i=3}^8 C_i \O_i$.
The $sdg$ operator $O_8$ which is relevant to 
${\rm Re} (\epsilon^{'}/\epsilon_K )$ is defined as 
\begin{equation}
\O_8 = {g_s \over 4 \pi}  m_s \ol{d}_L^\al \si^{\mu\nu} T^a s_R^\al 
                G^a_{\mu\nu}, 
\end{equation}
and other four quark operators $O_{i=3,...,6}$ and the corresponding Wilson 
coefficients from $C_3$ to $C_8$ with a single mass
insertion are available in the literature~\cite{gabbiani:96}. 
One has to remind that $C_{3,...,6}$'s are proportional to 
$(\delta_{12}^d )_{LL}$, whereas $C_8$ is given by a linear combination of
$(\delta_{12}^d )_{LL}$ and $(\delta_{12}^d )_{LR}$. This 
$(\delta_{12}^d )_{LR}$ dependent part in $C_8$ is proportional to 
$m_{\tilde{g}}/ m_s$, and thus is very important for generating
${\rm Re} (\epsilon^{'} / \epsilon_K)$ even if 
$(\delta_{12}^d )_{LR}$ is fairly small.

\begin{figure}[h]
\centerline{\epsfxsize=5.0cm \epsfbox{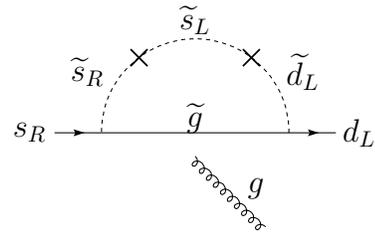}}
\caption{ Feynman diagram for $\Delta S=1$ process. The cross denotes the 
flavor changing $(LL)$ and the flavor preserving $(LR)$ mixings, 
respectively.}
\label{fig1}
\end{figure}

If we consider the penguin diagram Fig.~\ref{fig1} with the double mass 
insertion, the Wilson coefficient $C_8$ is given by 
\begin{equation}
\label{newdel1}
   C_8^{(2)} = {\al_s \over \wt{m}^2} \; {m_{\wt{g}} \over m_s} 
           (\de_{12}^d)_{LR}^{\rm ind} 
            \l[ \C_1 M_1^{(2)}(x) +\C_2 M_2^{(2)}(x) \r],
\end{equation}
where $\C_1=3/2, \C_2=-1/6$ and
\begin{eqnarray}
  M_1^{(2)}(x) &=& \frac{2(3-3x^2+(1+4x+x^2)\log x)}{(x-1)^5} \nl
  M_2^{(2)}(x) &=& \frac{1+9x-9x^2-x^3+6x(1+x)\log x}{(x-1)^5} .
\end{eqnarray}
The contributions of photon penguin and $Z$ penguin diagrams with the double 
mass insertion are negligible as in the case of the single mass insertion 
~\cite{gabbiani:96}.

Now we are ready to calculate the SUSY contributions to 
$\Delta m_K, \epsilon_K$ and $\epsilon' / \epsilon_K$ using the 
$\Delta S = 1,2$ effective Hamiltonians obtained above and the following 
expressions \cite{buras}:
\begin{eqnarray}
\Delta m_K ( {\rm SUSY} )  & = & 2 {\rm Re} M_{12} ,
\nonumber  
\\
\epsilon_K ({\rm SUSY})    & = & 
\frac{\mbox{exp}(i\pi/4)}{\sqrt{2}\Delta m_K ({\rm exp})}~{\rm Im} M_{12},
\\
{\rm Re} ( \epsilon' / \epsilon_K )  & = & 
\frac{\omega}{\sqrt{2} |\epsilon_K| {\rm Re} A_0}
\sum_{i} {\rm Im} ( C_i) \langle O_i \rangle_0 ( 1- \Omega_{\eta+\eta'})
\nonumber
\end{eqnarray} 
where $2 m_K M_{12}^* \equiv \langle \bar{K}^0 | H_{eff}^{\Delta S=2} 
| K^0 \rangle$ and $A_I$'s are the isospin amplitudes defined as
$A_I e^{i \delta_I} \equiv \langle ( \pi\pi)_I | {\cal H}_{\rm eff}^{\Delta
S = 1} | K^0 \rangle$.  
In the numerical analysis, we use the same parameters as in Ref.~\cite{buras}
with $m_s (2 {\rm GeV}) = 130 {\rm MeV}$. 
The corresponding SM prediction for ${\rm Re} ( \epsilon^{'} / \epsilon_K )
= 5.7 \times 10^{-4}$.  We vary the modulus and the phase of 
$( \delta_{12}^d )_{LL}$ as indepent parameters, and select those points 
which satisfy $\Delta m_K ( {\rm SUSY} ) \lesssim \Delta m_{K} ({\rm exp})$ 
and $| \epsilon_K ({\rm SUSY}) - \epsilon_K ({\rm exp}) | < 1 \sigma $. 
Then, for these points, we plot $\epsilon' / \epsilon_K$ in Figs.~2 (a)--(d) 
as functions of the modulus $r$ [(a) and (c)] and the phase $\varphi$ 
[(b) and (d)] of the parameter $(\delta_{12}^d)_{LL} \equiv 
r e^{i \varphi}$  for the common squark mass $\tilde{m}
= 500$ GeV. The upper (lower) rows correspond to $\wt{A_s} \equiv 
(A_s - \mu^* \tan\beta ) = -10~(-20)$ TeV. Different $x$'s (= $0.3,~1.0,~2.0$) 
are represented by the solid, the dashed and the dotted curves, respectively.
If we choose the opposite sign for $\wt{A_s}$, the phase of the relevant 
$(\delta_{12}^d )_{LL}$ should be shifted by 180$^{\circ}$ in order that we
have correct sign for $\epsilon_K$.  From Figs.~2 (a) and (b), it is clear 
that both $\epsilon_K$ and ${\rm Re} (\epsilon' / \epsilon_K)$ can be nicely 
accommodated with a single complex number $(\delta_{12}^{d})_{LL}$ with 
$\sim O(1)$ phase in our 
model without any difficulty, if $| \mu| $ and $\tan\beta$ is relatively 
large so that $|\wt{A_s}| $ becomes a few tens of TeV.  If the common squark 
mass $\tilde{m}$ differs from 500 GeV, the $\wt{A_t}$ should be multiplied by
$(\tilde{m} {\rm ~in ~GeV} / 500 )^2$ for the fixed $x$. 

Let us consider the neutron EDM constraint. 
The FP $LR$ mass insertion in the gluino-squark diagram 
contributes to the neutron EDM. 
The effective Hamiltonian for the neutron EDM is given by \cite{brhlik:98}
${\cal H}_{eff} ({\rm EDM}) =\sum_{i=1}^{3} C_{i}^{\rm edm} O_{i}, $
where 
the $O_i$'s are defined as 
\begin{eqnarray}
O_{1}&=& -\frac{i}{2} \bar{f} \sigma_{\mu\nu} \gamma_5 f 
F_{\mu\nu}, \nonumber \\
O_{2}&=& -\frac{i}{2} \bar{f} \sigma_{\mu\nu} \gamma_5 T^a f 
G^a_{\mu\nu}, \nonumber   \\ 
O_3&=& -\frac{1}{6} f_{abc} G^a_{\mu\rho} G^{b \rho}_{\ \nu} 
G^c_{\lambda\sigma}
\epsilon^{\mu\nu\lambda\sigma}. 
\end{eqnarray}
In our model, the FP $LR$ mass insertion in the gluino-squark diagram 
contributes to the neutron EDM with the following Wilson coefficients : 
\bea
C_1^{\rm edm} &=&- {2 \over 3} \; {e \al_s \over \pi} \; Q_d \;
                      {m_{\wt{g}} \over \wt{m}^2} \;
                      {\rm Im} \l(\de^d_{11} \r)_{LR} \;
                      B^{(1)}(x), \nl
C_2^{\rm edm} &=& {g_s \al_s \over 4 \pi} \;
                      {m_{\wt{g}} \over \wt{m}^2} \;
                      {\rm Im} \l(\de^d_{11} \r)_{LR}\;
                      C^{(1)}(x),
\eea
where $( \de^d_{11} )_{LR} \equiv m_d (A_d^* -\mu \tan \be)/\wt{m}^2$, and
\bea
  B^{(1)}(x) &=&
   {\frac{1 + 4\,x - 5\,{x^2} + 2\,x\,\left( 2 + x \right) \,\log (x)}
      {2\,{{\left( -1 + x \right) }^4}}}, 
\\
  C^{(1)}(x) &=&
   {\frac{2\,\left( -11 + 10\,x + {x^2} \right)  - 
        \left( 9 + 16\,x - {x^2} \right) \,\log (x)}{3\,
             {{\left( -1 + x \right) }^4}}}. \nonumber 
\eea
and $C_3^{\rm edm} = 0$. 
Our expression for $C_1^{\rm edm}$ confirms the result obtained in Ref.~
\cite{gabbiani:96}, and the result for $C_2^{\rm edm}$ is new.  
The renormalization group (RG) running effect and the final formula
for the neutron EDM can be found in~\cite{brhlik:98}.  We found that the EDM 
constraint is very strong. If the universality of the trilinear couplings are 
assumed ($A_d = A_s$), then $\l(\de^d_{11} \r)_{LR}$ should be essentially 
real. Still a single CP violating phase in $( \delta_{12}^d )_{LL}$ can 
generate right amounts of $\epsilon_K$ and ${\rm Re} (\epsilon' / \epsilon_K)$
without any fine tuning.  
Even if we relax the universality condition $A_d =  A_s$, the result is 
basically the same, since we are in the regime of large $\mu \tan\beta$ and
its phase is constrained by the neutron EDM irrespective of $A_d = A_s$ as 
long  as $| A_{d,s} | \lesssim 1$ TeV. Note that we are not assuming any 
specific flavor structures in the $A$ terms at all, unlike many other models
in Ref.~\cite{recent}.


It should be worthwhile to emphasize the importance of the FP
$(LR)$ mixing $(\delta_{22}^d )_{LR}$ in our study. If we ignored this effect 
and considered the flavor changing $(LL)$ and $(RR)$ mixings simultaneously 
for example, we could get both $\epsilon_K$ and ${\rm Re} (\epsilon' / 
\epsilon_K)$, but significant amounts of  fine tunings are unavoidable.
The ratio of the magnitues of $(\delta_{12}^d )_{LL}$ and 
$(\delta_{12}^d )_{RR}$ should be ${\cal O}(10^{-3})$ in order that we 
explain the large experimental data for ${\rm Re} (\epsilon'/\epsilon_K)$.
Then, the contributions from $(LL)^2$ [ or $(RR)^2$, whichever the 
larger one ] and $(LL) \times (RR)$ terms should cancel with each other 
within a part in $10^3$ in order to reproduce the experimental value for 
$\epsilon_K$, thus requiring substantial fine tuning \cite{future}. 

If the simplifying assumption of the real CKM matrix is relaxed, there will 
be additional constributions to $\epsilon_K$ and ${\rm Re} (\epsilon^{'} / 
\epsilon_K)$ from the  SM and other SUSY loop diagrams.
If we assume that only the gluino-squark contribution considered above is 
comparable with the SM contribution, it would be possible that the 
$\epsilon_K$ is mainly dominated by the KM phase contributions, but the 
${\rm Re} (\epsilon^{'} / \epsilon_K)$ has  significant contributions from 
the induced $(\delta_{12}^d )_{LR}$ as disucssed in this letter 
\cite{future}. 

The best discriminant between our model and the SM model would be probably
the branching ratios for $K\rightarrow \pi \nu\nu$ and CP violations in 
$B$ decays. 
The branching ratio for the decay $K_L \rightarrow \pi^0 \nu\bar{\nu}$ is 
essentially zero in our model, since it is purely CP violating but there is 
no appreciable CP violation in the $s \rightarrow d \nu \bar{\nu}$ amplitude 
through gluino loop diagram \cite{gabbiani:96}.  On the contrary, 
$K^+ \rightarrow \pi^+ \nu\bar{\nu}$ involves both CP conserving and CP 
violating amplitudes, and the corresponding branching ratio in our model 
ranges over $(8.9 \pm 7.2) \times 10^{-11}$, compared to the SM prediction : 
$B ( K^+ \rightarrow \pi^+ \nu\bar{\nu} )_{\rm SM} = (7.7 \pm 3.0) \times 
10^{-11}$ \cite{kpinn}. Our predictions can be changed by two ways, however. 
The chargino-upsquark loop contributions contribute to 
$K\rightarrow \pi \nu\bar{\nu}$ through the enhanced $sdZ$ penguin vertex, 
but one still expects the branching ratio for 
$K_L \rightarrow \pi^0 \nu\bar{\nu}$ to be smaller than the SM predictions 
\cite{kpinn1}. 
Also, if our assumption on the real CKM matrix is relaxed, the KM phase will
contribute to the $K\rightarrow \pi \nu\bar{\nu}$, but the predictions
will differ from the SM, since there is additional contribution to 
$\epsilon_K$ so that the CKM elements are less constrained \cite{future}.  
If we stick to $\delta_{KM} =0$, the CP violations in $B$ system will be very
different from the SM predictions. However, the gluino-squark loop 
contributions to the $B$ system is governed by new parameters, 
$(\delta_{13}^d)_{AB}$ and $(\delta_{23}^d)_{AB}$, which are independent of 
$(\delta_{12}^d)_{AB}$ we considered 
here. Therefore we cannot make defninte predictions for $B$ decays.  
Generically the situation could be very different from the SM case \cite{kkl}.

Now let us consider a typical size of $(\delta_{12}^d )_{LL}$ in the MSSM.
The answer to this question would depend on the models for soft SUSY breaking
and how to solve the SUSY flavor problem. If one invokes approximate (abelian
or nonabelian) flavor symmetry in order to solve the SUSY FCNC problem, 
the natural size of $(\delta_{12}^d )_{LL}$ could be order of 
$\sim (\lambda^3 -\lambda^4) \sim 10^{-3}$ with order $O(1)$ phase 
(where $\lambda =\sin\theta_c =0.22$)~\cite{nir}. This is usually referred
as the alignment in contrast to more popular universality. 
This is in fact the twisted version of the so-called SUSY FCNC and SUSY
$\epsilon_K$ problem, saying that the gluionic SUSY contribution to 
$\epsilon_K$ could be generically very large unless the squarks are degenerate
and/or the mass matrices of quarks and squarks are almost aligned in the 
flavor space. In this context, our result on $\epsilon^{'}/\epsilon_K$ is 
nothing but to say that SUSY $\epsilon_K$ problem in general implies the SUSY 
$\epsilon^{'}$ problem for relatively large 
$| \mu \tan\beta | \sim O(10)$ TeV.

In conclusion, we showed that both $\epsilon_K$ and 
${\rm Re} (\epsilon' / \epsilon_K )$ 
can be accommodated with a single CP violating and flavor changing 
down-squark mass matrix elements $[ (\delta_{12}^d )_{LL} \sim 10^{-3} ]$ 
without any fine tuning or any conflict with the data on FCNC processes,
if $| \mu \tan\beta| \sim 10-20$ TeV with a scale factor 
$[ \tilde{m} ({\rm in~ GeV}) / 500 ]^2$ for the fixed $x$. 
Our mechanism utilizes this FC $LL$ mass insertion along with the FP $LR$ 
mass insertion propotional to $(\delta_{22}^d )_{LR} \sim 10^{-2}$.  
The latter is generically present in any SUSY models including the MSSM, 
and thus there is no fine tuning in our model for accommodating both 
$\epsilon_K$ and ${\rm Re} (\epsilon' / \epsilon_K)$ in terms of a single
$(\delta_{12}^d )_{LL} $. 
It is straightforward to extend our mechanism including the nonvanishing 
KM phase, $( \delta_{12}^d )_{LR}$ and/or $( \delta_{12}^d )_{RR}$. 
One can also consider our mechanism in the more minimal SUSY model, where 
$(\delta_{22})^d_{LR}$ is proportional to $m_b ( A_b - \mu \tan\beta)$ 
so that $A_b - \mu \tan\beta$ may be lowered significantly.
All these finer details including 
$B( K \rightarrow \pi \nu \bar{\nu} )$ will be discussed 
elsewhere in the forthcoming publication \cite{future}. 

\acknowledgements
P.K. is grateful to Kiwoon Choi and S. Pokorski for useful discussions on 
the subject presented here.
This work was supported by BK21 project of the Ministry of Education, 
and by KOSEF Postdoctoral Fellowship Program (SB).



\psfrag{XX1}[][][0.85]{{\Large $s_R$}}
\psfrag{XX2}[][][0.85]{{\Large $\wt{g}$}}
\psfrag{XX3}[][][0.85]{{\Large $d_L$}}
\psfrag{XX4}[][][0.85]{{\Large $\wt{s}_R$}}
\psfrag{XX5}[][][0.85]{{\Large $\wt{s}_L$}}
\psfrag{XX6}[][][0.85]{{\Large $\wt{d}_L$}}
\psfrag{XX7}[][][0.85]{{\Large $g$}}

\begin{figure}[h]
\centerline{\epsfxsize=9.3cm \epsfbox{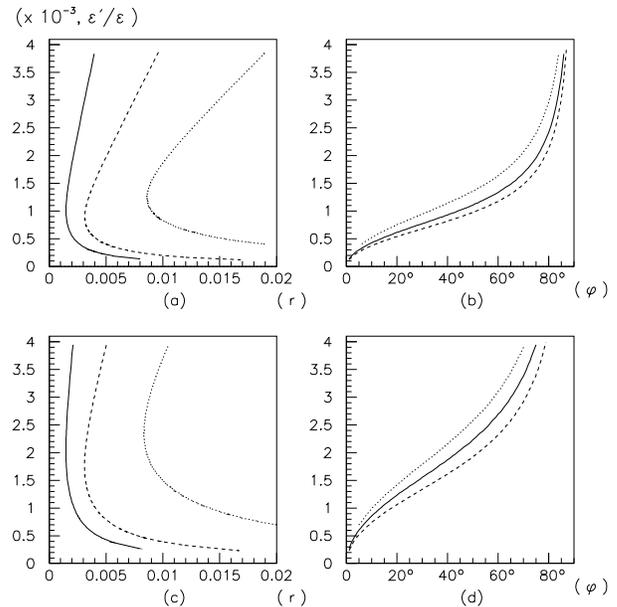}}
\caption{ $Re ( \epsilon' / \epsilon_K )$ as a function of the modulus $r$
[(a) and (c)] and the phase $\varphi$ [(b) and (d)] of the parameter 
$(\delta_{LL}^d )_{12}$ 
with $\wt{A_S}$ to be $-10~TeV$ ((a),(b)) 
and $-20~TeV$ ((c),(d)).  
The common squark mass is chosen to be $\wt{m} = 500$ GeV, and the solid, 
the  dashed and the dotted curves correspond to $x = 0.3, ~1.0, ~2.0$, 
respectively.
}
\label{fig2}
\end{figure}


\end{multicols}
\vfil\eject
\end{document}